\documentclass{iauc}
\usepackage{graphicx}

\title[Structural properties of dEs and Es]{Structural properties of dwarf 
ellipticals and the connection with (ordinary) elliptical galaxies}

\author[Alister W.\ Graham]
{Alister W.\ Graham}

\affiliation{Mount Stromlo and Siding Spring Observatories,
Australian National University, Private Bag, Weston Creek PO, ACT
2611, Australia.
\break 
email: Graham@mso.anu.edu.au}

\pubyear{2005}
\volume{198} 
\pagerange{001--008}
\date{?? and in revised form ??}
\jname{Near--Field Cosmology with Dwarf Elliptical Galaxies}
\editors{H.\ Jerjen and B.\ Binggeli, eds.}
\begin{document}

\maketitle

\begin{abstract}
This article reviews the popular reasons for the belief that dwarf
elliptical galaxies and (ordinary) elliptical galaxies are distinct
and separate species.  They include: light--profile shape (or
similarly image concentration); the magnitude--central surface
brightness diagram; the magnitude--effective surface brightness
diagram (or similarly the magnitude--effective radius diagram); and
the Fundamental Plane.  It is shown how a continuous trend between
luminosity and a) light--profile shape, and b) central surface
brightness (until the onest of core formation at $M_B \sim -20.5$
mag), results in a unification of the dwarf elliptical and (ordinary)
elliptical galaxies.  Neither the above four reasons, nor the
luminosity function (at least in the Virgo cluster) provide evidence
for a division at $M_B=-18$ mag between the dwarf elliptical and
(ordinary) elliptical galaxies.  Instead, they appear to be continuous
extensions of each other.
%
\end{abstract}

\firstsection 
\section{Introduction}

For at least the past two decades, astronomers have believed that
dwarf elliptical (dE) galaxies (not to be confused with dwarf
spheroidal galaxies, which typically have absolute magnitudes fainter
than $M_B=-13$ mag) are distinct objects from the more luminous
(ordinary) elliptical (E) galaxies.  The divide between these
allegedly different classes is supposed to occur at an absolute
magnitude of $M_B = -18$ mag (e.g., Kormendy 1985).

In this talk I will review the most oftenly quoted reasons why the dEs
are thought to be different from the Es, and then go on to explain why
I think these past claims for a discontinuity at $M_B = -18$ mag
should be reconsidered.  I will argue that a continuity in structural
properties exists across this alleged boundary.

Much, but not all, of this talk is based on Sections 1 and 4 from
Graham \& Guzm\'an (2003), where further details may be found.

\section{Past evidence of continuity}

Before addressing the issue of galaxy structure, which is where 
past claims of discontinuity have their roots, it is relevant to
review various other physical aspects of the dE and E population.
Several physical quantities vary smoothly across the alleged division
at $-$18 $B$--mag. 

The average globular cluster metallicity is known to vary
continuously, as a function of host galaxy magnitude, across the
alleged dE/E divide (e.g., Forbes et al.\ 1996, their
figure 14).  C\^ot\'e (2005, these proceedings) has shown how this arises
from a systematic variation in the relative numbers of red and blue
globular clusters, as opposed to a single population changing its mean
metallicity.

Furthermore, the host galaxies themselves follow a
luminosity--metallicity relation that shows no sign of a division or
break at $M_B = -18$ mag (e.g., Caldwell \& Bothun 1987).  Related to
this, the colour--magnitude relation shows no sign of a division at
$M_B = -18$ mag (e.g., Caldwell 1983). 
Now, on average, dEs do have a different metallicity from the
brighter Es, but the key point is that there is a continuous change in
chemical composition between the two. 
If one was to measure the average height and weight of a sample of
adults, and then do the same for a sample of children, one would of
course conclude that these populations have different physical
properties.  But this obviously tells us little as to whether or 
not children are the same specie as adults. 

Aside from the stellar populations not knowing about the supposed dE/E
transition, there is also no dynamical evidence.  Including dozens of
dEs, Matkovi\'c \& Guzm\'an (2005, and also these proceedings) and De
Rijcke et al.\ (2005, see also these proceedings) have both shown that
the luminosity--velocity dispersion ($L$--$\sigma$) relation continues
linearly across the alleged boundary between the dEs and Es.
Interestingly, both studies found that the slope of this relation is
$\sim$2, jumping to $\sim$4 and thus recovering the Faber--Jackson
(1976) relation at magnitudes greater than $\sim -20.5$ $B$--mag.


What then, are the popular reasons which have led people to believe
that the dEs form a uniquely different class of object to the Es?

\section{Past evidence of discontinuity} 

This section lists the three most often quoted reasons why
dwarf and giant elliptical galaxies are thought to be different
species of object.  The reasons pertain to galaxy structure. 
In the light of new data, however, the following section will 
explain why these three arguments are no longer valid.  The following
section will also counter two further claims of support for a dE/E 
discontinuity at $M_B = -18$ mag. 

{\it (i)} Dwarf elliptical galaxies have exponential light--profiles
(Faber \& Lin 1983; Binggeli, Sandage \& Tarenghi 1984), while
elliptical galaxies have $R^{1/4}$ profiles (de Vaucouleurs 1948,
1959; de Vaucouleurs \& Capaccioli 1979).

{\it (ii)} Dwarf elliptical galaxies define a distribution that is
almost at right angles to that traced by the brighter elliptical
galaxies in the diagram of central surface brightness, $\mu_0$, versus
absolute magnitude, $M$, (Kormendy 1985, his figure 3).  Such behaviour
has been taken as evidence that different formation processes must
have shaped the dEs and the Es.  The divide is said to occur at an
absolute magnitude of $-$18 $B$--mag.

{\it (iii)} In diagrams which plot a galaxy's effective surface
brightness, $\mu_{\rm e}$, dwarf elliptical galaxies follow a
distribution that is clearly offset to that defined by the brighter
elliptical galaxies.  This is the case for the $M$--$\mu_{\rm e}$ and
$R_{\rm e}$--$\mu_{\rm e}$ diagram (Capaccioli, Caon \& D'Onofrio
1992, their figure 4; Binggeli \& Ferguson 1994; their section 2.2.2),
and also for the Fundamental Plane (Djorgovski \& Davis 1987) when
constructed using effective surface brightnesses (e.g., Geha,
Guhathakurta \& van der Marel 2002).

\section{A unified picture}

In the Author's opinion, there have been two key advances since the
above three observations were established.  The first has come from a
closer scrutiny of optical light--profiles.  Studies such as those by
Schombert (1986) revealed that the $R^{1/4}$--law was not as universal
as previously thought.  Similarly, analyses by Caldwell \& Bothun
(1987) and Binggeli \& Cameron (1991) revealed that the dEs have 
magnitude--dependent departures from the exponential model.  The
subsequent use of high--quality CCD imagery no doubt helped to refine
this picture.  Second, the superb resolution provided by the Hubble
Space Telescope has allowed us to probe the centers of galaxies on
scales previously washed out by atmospheric seeing.  The impact of
these two things on the above three points is given below.

\begin{figure}
\begin{centering}
\includegraphics[height=12cm,angle=270]{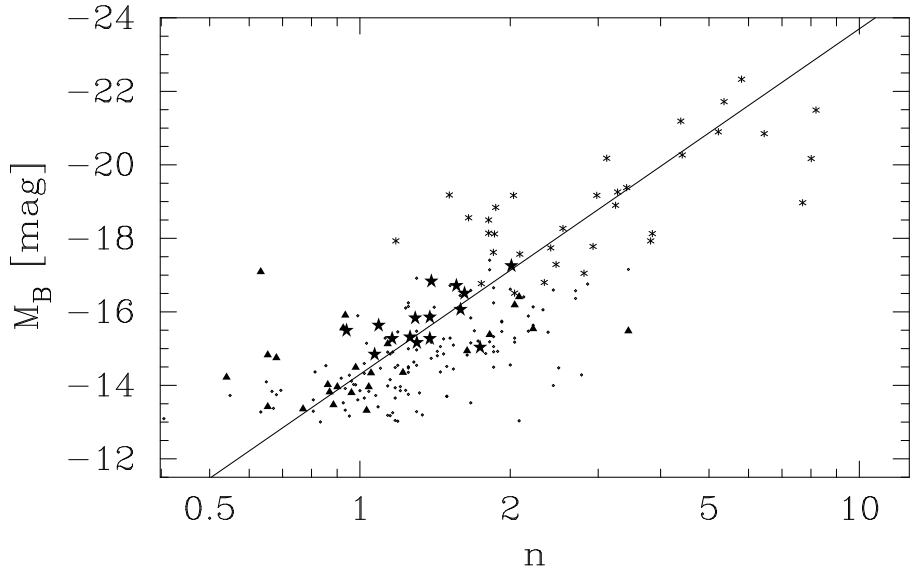}
\caption{
Absolute $B$--band galaxy magnitude versus the logarithm of the
S\'ersic shape index $n$.  Due to biasing from the magnitude cutoff at
$M_B\sim -13$ mag, the line $M_B = -9.4\log(n) - 14.3$ has been
estimated by eye.  Dots represent dE galaxies from Binggeli \& Jerjen
(1998), triangles are dE galaxies from Stiavelli et al.\ (2001), large
stars are dE galaxies from Graham \& Guzm\'an (2003), asterix are
intermediate to bright E galaxies from Caon et al.\ (1993) and
D'Onofrio et al.\ (1994).  Lenticular galaxies have been excluded.
}
\label{Fig_M-n}
\end{centering}
\end{figure}

{\it (i)} There is no question that de Vaucouleurs' $R^{1/4}$ model
fits many elliptical galaxies remarkably well. It is a clever and
useful empirical function.  However, as noted in Djorgovski \&
Kormendy (1989), it fits best elliptical galaxies with $M_B \sim -21$
mag; brighter and fainter galaxies having different curvature than
described by the $R^{1/4}$ model.  Thanks to the pioneering work of
Capaccioli and collaborators, most notably Caon, Capaccioli \&
D'Onofrio (1993), numerous studies have now shown that S\'ersic's
(1963, 1968) $R^{1/n}$ generalisation\footnote{A useful compilation of
various S\'ersic expressions can be found in Graham \& Driver (2005).}
accommodates for the real (Trujillo, Graham \& Caon 2001; their
section 2) and systematic changes in light--profile shape $n$ with
absolute magnitude (e.g., D'Onofrio, Capaccioli, \& Caon 1994; Graham
et al.\ 1996; Vennik et al.\ 1996).  Similarly, the stellar
distribution in dwarf elliptical galaxies is also a function of
absolute magnitude (e.g., Davies et al.\ 1988; Cellone, Forte, \&
Geisler 1994; Vennik \& Richter 1994; Young \& Currie 1994, 1995;
Karachentseva et al.\ 1996; Jerjen \& Binggeli 1997).  This is shown
in Figure~\ref{Fig_M-n} using a heterogeneous sample of literature
data.  Presumably the actual level of scatter is less than seen here.
One may alternatively plot concentration, rather than profile shape,
as done in Graham, Trujillo \& Caon (2001).

\begin{figure}
\begin{centering}
 \includegraphics[width=12cm,angle=0]{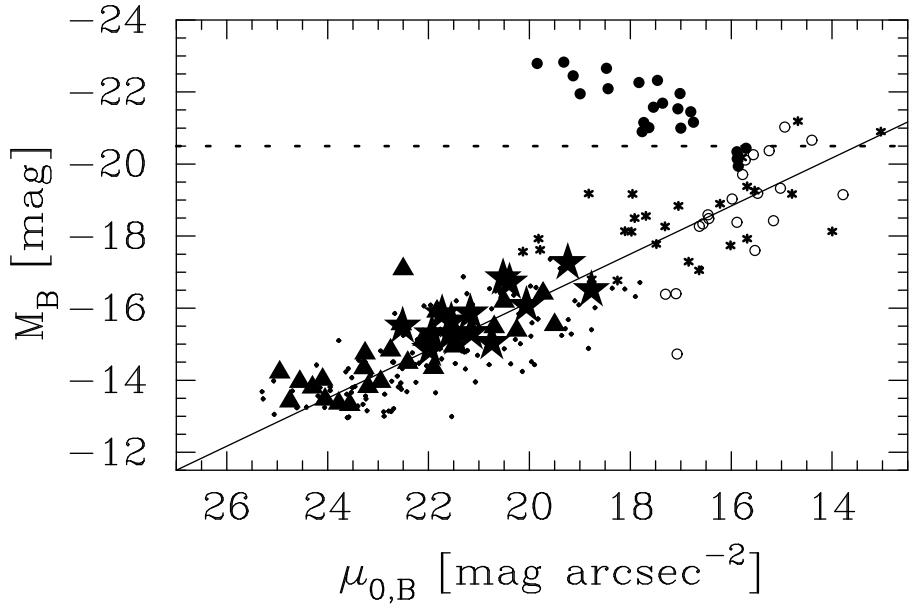}
  \caption{
Galaxy magnitude versus (host galaxy) central surface brightness.  Due
to biasing from the magnitude cutoff at $M_B\sim -13$ mag, the line $M_B =
(2/3)\mu_0 - 29.5$ has been estimated by eye rather than from a linear
regression routine.  The central surface brightness values obtained
from S\'ersic models fitted to luminous ($M_B\lesssim -20.5$ mag) E
galaxies follow this same relation (Jerjen \& Binggeli 1997), although
they were not used to define it. The symbols are the same as in
Figure~\ref{Fig_M-n}, with the additional open circles representing
the so--called ``power--law'' E galaxies from Faber et al.\ (1997),
and the filled circles representing the ``core'' E galaxies from these
same authors.
}
\label{Fig_M-mu0}
\end{centering}
\end{figure}

{\it (ii)} Using {\sl HST} data from the Nuker team (Faber et al.\
1997), the $M$--$\mu_0$ relation is now much better defined than it
was twenty years ago, especially at the bright end.  Furthermore, the
increased number of galaxies around the alleged transition magnitude
of $-$18 $B$--mag has also improved our understanding of what is going
on.
%
%
What is immediately clear from Figure~\ref{Fig_M-mu0} is that there is
no break at $M_B = -18$ mag.  Instead, a linear trend exists across
this supposed divide, and extends from $M_B=-13$ mag to $M_B \sim
-20.5$ mag.  This diagram does not imply that the dEs ($M_B > -18$
mag) experienced a different formation process to the Es.

The behaviour of the elliptical galaxies in Figure~\ref{Fig_M-mu0} is
similar to that seen in the $L$--$\sigma$ diagram, in which galaxies
more luminous than $M_B \sim -20.5$ mag depart from the linear
distribution defined by the less luminous Es and dEs (see Matkovi\'c 
\& Guzm\'an 2005, these proceedings, their figure 1).

The deviant galaxies at the high--luminosity end in
Figure~\ref{Fig_M-mu0} have all been identified by the Nuker team as
``core'' galaxies: their central light--profiles display a clear break
and downward flattening relative to the inward extrapolation of their
outer (S\'ersic) stellar distribution (Graham et al.\ 2003; Trujillo
et al.\ 2004).
Such ``core'' galaxies may have formed from the
dissipationless merger of two or more lesser Es, in which the
progenitor galaxy's supermassive black holes have coalesced at the
center of the new galaxy via the gravitational slingshot, and thus
evacuation, of the central stars (Begelman, Blandford, \& Rees 1980;
Ebisuzaki, Makino, \& Okumura 1991).  This local, rather than global,
physical process, reduces the central flux by $\sim$0.1\% of the total
galaxy flux (Graham 2004).

It is worth noting that the central surface brightness one would
obtain from the inward extrapolation of the undisturbed outer
light--profile of the ``core'' galaxies yields values that fall on the
linear relation shown in Figure~\ref{Fig_M-mu0} (Jerjen \& Binggeli
1997).

\begin{figure}
\begin{centering}
\includegraphics[width=6.3cm,angle=270]{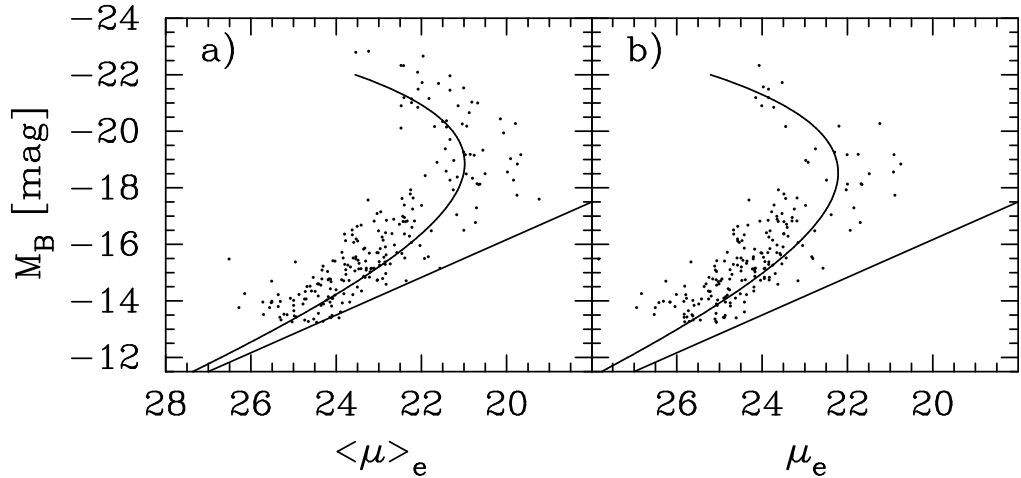}
\caption{
The curved lines show the expected behavior of galaxies in
the (a) magnitude--mean effective surface brightness and (b)
magnitude--effective surface brightness diagrams given the linear 
relations seen in Figures~\ref{Fig_M-n} and \ref{Fig_M-mu0}.  
The straight line shown here is from Figure~\ref{Fig_M-mu0}.
}
\label{M_mue}
\end{centering}
\end{figure}

{\it (iii)} The difference between the central surface brightness, 
$\mu_0$, and the effective surface brightness, $\mu_{\rm e}$ (the
surface brightness at the effective half--light radius $R_{\rm e}$), is
the same for every $R^{1/4}$ profile.  This difference,
however, varies as a function of profile shape $n$, and is 
given by the expression $\mu_{\rm e} = \mu_0 + 1.086b$, where $b \sim
(2n - 1/3)$.  So, as the galaxy magnitudes 
brighten in Figure~\ref{Fig_M-mu0}, the value of $n$ increases
(Figure~\ref{Fig_M-n}) and thus the value of $\mu_{\rm e} - \mu_0$
increases.  This is shown in Figure~\ref{M_mue}b.  Related to this,
the mean surface brightness within the effective radius, 
$\langle\mu\rangle_{\rm e}$, can be derived from the expression
$\langle\mu\rangle_{\rm e} = \mu_{\rm e} -
2.5\log[e^{b}n\Gamma(2n)/b^{2n}]$, where $\Gamma$ is the gamma
function (e.g., Graham \& Driver 2005).

The different slopes for the dE and E galaxy distributions in
Figure~\ref{M_mue} is merely a consequence of a continuously
varying profile shape with galaxy luminosity, it does not imply
distinctly different galaxy formation processes for dEs and Es.

\subsection{Other issues}

{\it (a)} Kormendy (2004) has claimed that the decline in the 
dE galaxy luminosity function at bright magnitudes, and the 
associated decline in the 
E galaxy luminosity function at faint magnitudes is evidence 
that they are distinct populations.  This argument of course 
falls down if one assigns a galaxy to the dE or E population 
based on its luminosity.  Similarly, if one assigns an object to 
a class based on its appearance (i.e., using image concentration or 
light--profile shape), the situation remains unchanged 
(see Figure~\ref{Fig_M-n}). 

\begin{figure}
\begin{centering}
\includegraphics[height=8.0cm]{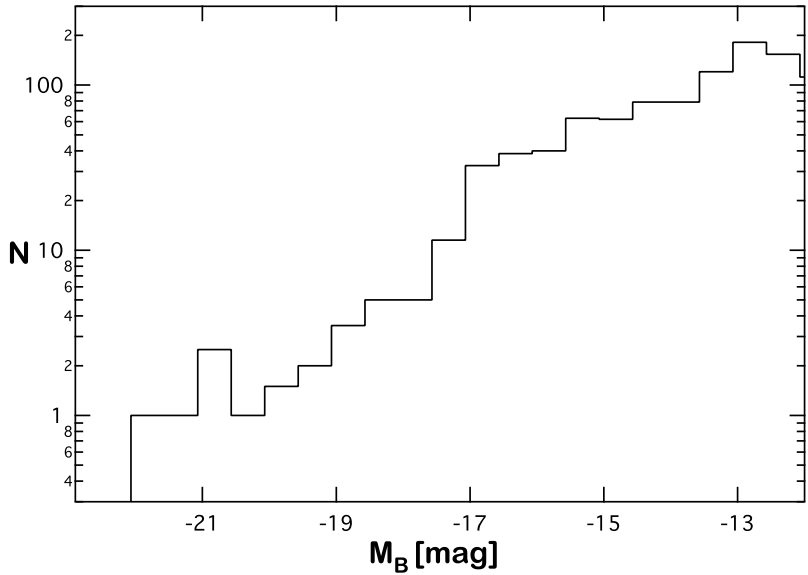} 
\caption{
Virgo cluster dE $+$ E galaxy luminosity function.  There is no
evidence for a division at $M_B = -$18 mag.  Likely (50\% chance)
cluster members are weighted by 0.5.  Figure courtesy of H.\ Jerjen. 
}
\label{cmon}
\end{centering}
\end{figure}

Figure~\ref{cmon} shows the luminosity function of the Virgo dE $+$ E
galaxies.  There is obviously nothing special about $M_B = -18$ mag,
and certainly no evidence for a deficit of galaxies at this magnitude,
nor is it the overlap region of two distributions peaking at fainter
and brighter magnitudes.

{\it (b)} From the beginning it has been known that galaxies with 
central velocity dispersions less than about 100 km s$^{-1}$ do not
follow the ``Fundamental Plane'' (Djorgovski \& Davis 1987) as 
defined by luminous elliptical galaxies. 

Using $L=2\pi R_{\rm e}^2 \langle I \rangle_{\rm e}$, it can be shown
that the same mechanism that causes the curved relations in
Figure~\ref{M_mue} is also behind the different slopes in the
$\langle\mu\rangle_{\rm e}$--$\log R_{\rm e}$ and $M-\log R_{\rm e}$
diagram for the dEs and Es.  It therefore comes as no surprise that
the dEs are not distributed on the Fundamental Plane.  As was noted in
Graham \& Guzm\'an (2004), the Fundamental Plane is simply the tangent
sheet to the high--luminosity end of a curved surface in the
three--space of $\langle\mu\rangle_{\rm e}, R_{\rm e},$ and $\sigma$.
Only bright Es that follow the Kormendy (1977) relation (see
Capaccioli et al.\ 1992, their figure 4) will reside on typical
constructions of the Fundamental Plane.  The use of central surface
brightness mitigates the apparent different behaviour of the dEs and
Es (Guzm\'an et al. 2005, in prep).

\section{Conclusions}

Chemically and dynamically, the dEs appear to form a continuous
extension with the Es.  Both the colour--magnitude relation 
(but see Lisker et al., these proceedings) 
and the 
$L$--$\sigma$ relation show a continuous linear trend across the
alleged boundary at $-$18 $B$--mag.  Similarly, the dE $+$ E galaxy
luminosity function shows no evidence of a divide at this magnitude.

In terms of galaxy structure, dEs and Es display a continuous range of
properties that vary smoothly across the alleged dE/E divide.  
The shape of their (projected) stellar distribution,
quantified through S\'ersic's index $n$, increases steadily with
luminosity from values around 0.5 for the fainter dwarfs, to values
greater than 4 for the more luminous ellipticals.
The central (host galaxy) surface brightness brightens linearly (from
$M_B \sim -13$ mag) as the host galaxy magnitude brightens.  There is
no discontinuity at $M_B = -18$ mag in the $M$--$\mu_0$ diagram. 

The above two relations account for the near--orthogonal distribution
of dEs and Es in diagrams of effective surface brightness versus (i)
magnitude and (ii) effective radius.  They also explain why the dEs
depart from the Es in the traditional construction of the Fundamental
Plane, which uses effective surface brightness.

We report a clear break in the $M$--$\mu_0$ relation at $M_B \sim
-20.5$ mag, such that more luminous galaxies have fainter central
surface brightnesses.  Thanks to the resolution of {\sl HST}, this
departure of the brightest elliptical galaxies from the main
$M$--$\mu_0$ relation is understood to be associated with the
presence of partially depleted cores.

\begin{acknowledgments}
I wish to thank my collaborators Rafael Guzm\'an, Ana Matkovi\'c and
Ileana Vass who have contributed to various aspects of this paper.  I
am also grateful to Helmut Jerjen who provided the luminosity function
shown in Figure~\ref{cmon}.
This research was supported in part by NASA grant HST-AR-09927.01-A
from the Space Telescope Science Institute, which is operated by the
Association of Universities for Research in Astronomy, Inc., under
NASA contract NAS5-26555.
\end{acknowledgments}

\begin{discussion}

\discuss{Taylor}{ What are the smallest black holes you need in
this picture to scour out the cored galaxies?}

\discuss{Graham}{ Bearing in mind that it's not simply the black hole
masses, but whether the galaxy merger is dissipationless or not --- as
gas can facilitate black hole coalescence and also help repopulate the
loss cone --- the smallest black holes in galaxies with cores have
masses around $10^8 M_{\odot}$.}

\discuss{Conselice}{ The low--luminosity ellipticals also have
black holes, and probably also form through mergers.  If this is the
case, as seems likely, then why don't the low--luminosity ellipticals
also have depleted cores?  Does this suggest that giant and fainter
ellipticals really do have two formation methods?}

\discuss{Graham}{ A good question.  Giant elliptical galaxies with
partially depleted cores tend to be brighter than $\sim -20.5$
$B$--mag, while ``power--law'' galaxies tend to be fainter.  It has
been suggested that the ``core'' galaxies formed via a dissipationless
merger event while the ``power--law'' galaxies formed from a
relatively gas rich merger.  No division at $-$18 $B$--mag is apparent
between the ``power--law'' ellipticals and the dwarf ellipticals.  In
passing I note that the boxy/disky isophote divide also occurs at
$\sim -20.5$ $B$--mag, as does the change in slope of the
$L$--$\sigma$ relation (Matkovi\'c \& Guzm\'an and de Rijcke et al.\
these proceedings).}

\end{discussion}

\end{document}